\documentclass[prb,notitlepage,showpacs,amsmath,amssymb,groupedaddress,longbibliography]{revtex4-2}
\usepackage{latexsym}
\usepackage{graphicx}
\usepackage{color}
\usepackage{xspace}
\usepackage{soul}
\usepackage[bookmarks=false,linkcolor=blue,urlcolor=blue,colorlinks,citecolor=blue]{hyperref}

\renewcommand{\thefigure}{\arabic{figure}}
\setcitestyle{super,open={},close={}}

\makeatletter
\renewcommand\@biblabel[1]{#1.}
\makeatother

\newcommand{\beq}    {\begin{equation}}
\newcommand{\enq}    {\end{equation}}
\newcommand{\ceq}[1] {(\ref{#1})}

\newcommand{\nbse}   {${\rm NbSe_2}$\xspace}

\newcommand{\mgb}    {${\rm MgB_2}$\xspace}
\newcommand{\cdas}   {${\rm Cd_3As_2}$\xspace}

\begin{document}
\title{Leggett Modes in Dirac Semimetals}

\author{Joseph~J.~Cuozzo$^{1,2}$}
\author{W.~Yu$^{3}$}
\author{P.~Davids$^{3}$}
\author{T.M.~Nenoff$^{3}$}
\author{D.B.~Soh$^{2}$}
\author{Wei~Pan$^{2}$}
\author{Enrico~Rossi$^{1}$}
\affiliation{$^{1}$ Department of Physics, William \& Mary, Williamsburg, VA 23187, USA\\
$^2$ Sandia National Laboratories, Livermore, CA 94551, USA.\\
$^{3}$ Sandia National Laboratories, Albuquerque, New Mexico 87185, USA 
}

\date{\today}

\maketitle

{\bf 
In recent years experimentalists
have been able to clearly show that several materials, such as \mgb~\cite{Tsuda2001,Souma2003}, 
iron-based superconductors~\cite{Stewart2011}, monolayer \nbse~\cite{Ugeda2015,Xi2016a}, are multiband superconductors.
Superconducting pairing in multiple bands can give rise to novel and very interesting phenomena.
Leggett modes~\cite{Leggett1966} are exemplary of the unusual effects that can be present in multiband superconductors.
A Leggett mode describes the collective periodic oscillation of the relative phase between the phases of 
the superconducting condensates formed by electrons in different bands. It can be thought of as the mode arising 
from an inter-band Josephson effect. The experimental observation of Leggett modes
is challenging for several reasons: 
(i)   Multiband superconductors are rare;
(ii)  they describe charge fluctuations between bands and therefore are hard to probe directly;
(iii) their mass gap is often larger than the superconducting gaps and therefore
	  are strongly overdamped via relaxation processes into the quasiparticle continuum.
In this work we show that Leggett modes, and their frequency, can be detected unambigously in 
a.c. driven superconducting quantum interference devices (SQUIDs).
We then use the results to analyze the measurements of a SQUID based on
\cdas, an exemplar Dirac semimetal, in which superconductivity is induced by proximity to superconducting Al.
The experimental results show the theoretically predicted unique signatures of Leggett modes
and therefore allow us to conclude that a Leggett mode is present in the two-band superconducting
state of Dirac semimetal (DSM) \cdas.}

\begin{figure}[tb!]
\centering
\includegraphics[width=1.0\textwidth]{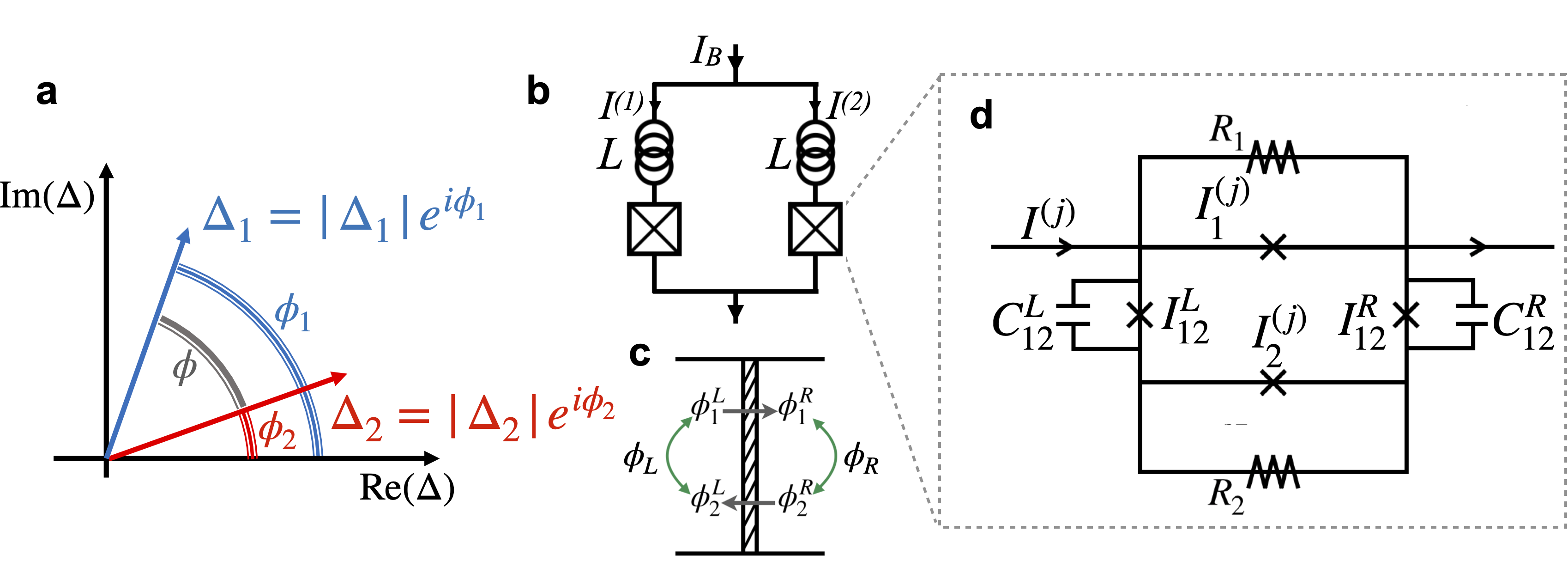}
\caption{\label{fig1}\textbf{Leggett modes in SQUIDs.} 
 \textbf{a} 
 Schematic showing the relative phase $\phi$  between the two superconducting order parameters.
 \textbf{b} 
 SQUID. The boxes represent indvidual JJs, whose effective RCJS model is shown in \textbf{d}.
 \textbf{c} Diagram showing the superconducting phases across a JJ.
 }
\end{figure}

Let us consider a system with two bands crossing
the Fermi energy. In the presence of superconducting pairing the ground state
of the system can be characterized by two complex superconducting order parameters 
$\Delta_i=|\Delta_i|e^{i\phi_i}$ for band $i=1$ and $i=2$, as shown schematically in Fig.~\ref{fig1}~(a).
In addition to collective modes associated to oscillations of the amplitude and phase of the 
individual order parameters, Leggett pointed out~\cite{Leggett1966} that an additional
collective mode describing oscillations of the relative phase $\phi=\phi_1-\phi_2$ could be present.
In the ideal case, the dynamics of the Leggett mode is described by the effective Lagrangian

\begin{equation}
 \mathcal{L} = (1/2)C_{12}(\hbar/2e)^2(d\phi/dt)^2 + (\hbar/2e)I_{12}\cos(\phi-\phi_0)
 \label{eq:L}
\end{equation}

where $C_{12}$ is the interband capacitance, $e$ is the electron's charge,
$I_{12}$ the effective interband critical Josephson current~\cite{Leggett1966},
and $\phi_0$ the equilibrium value of $\phi$.
From \ceq{eq:L} we obtain that when $\phi-\phi_0\ll 1$ $\phi$ will oscillate with frequency $\omega_L=\sqrt{(2e/\hbar)I_{12}/C_{12}}$ around $\phi_0$. 
The interband nature of the charge oscillations associated with the Leggett mode make its detection challenging.
So far, using spectroscopy techniques, signatures of a Leggett mode have been observed in \mgb~\cite{Brinkman2006,Blumberg2007,Klein2010,Mou2015,Giorgianni2019},
and, more recently, in a Fe-based superconductor~\cite{Zhao2020}.
Using an approach of limited applicability, it had been theorized that in Josephson junctions (JJs) in which one lead is formed by a single-band 
superconductor and the other by a two-band superconductor, signatures of a Leggett mode could be present~\cite{Ota2009z}.
Here, using a different method, we show that the presence of a Leggett mode can induce 
robust qualitative features in a SQUID in which all the leads are formed from the same multi-band superconducting material,
and verify experimentally the presence of such signatures in a SQUID based on the superconducting DSM \cdas.

A SQUID, see Fig.~\ref{fig1}~(b), is formed by two Josephson junctions connected in parallel and encircling a finite size area.
Let $\theta_{i}\equiv\phi_{iR}-\phi_{iL}$ be the difference between the superconducting order parameter in the right and left lead for band $i$.
We then have that the current across the JJ's leads, for a JJ with low-medium transparency~\cite{Beenakker1992a}, 
is given by $I=I_1\sin(\alpha_1\theta_1) + I_2\sin(\alpha_2\theta_2)$
where $I_i$ is the critical supercurrent for band $i$, and $\alpha_i$ is equal to 1 for a standard JJ, and
1/2 for a topological JJ~\cite{Fu2009d,Wiedenmann2016a,Dartiailh2021a}.
For biased high transparency topologically trivial JJs, Landau-Zener transitions 
can induce a current-voltage response equivalent to a topological junction~\cite{Dartiailh2021}.

To understand the effect of a Leggett mode on the dynamics, and voltage-current (V-I) characteristic, of a JJ,
we first present a simplified analysis of a voltage-biased JJ. A rigorous analysis of the realistic
case of a current-biased JJ is presented later, see also the SI~\cite{SI}.
In presence of a voltage $V$ across the JJ's
leads, the phases $\theta_i$ evolve over time according to the equations $d\theta_i/dt = 2eV/\hbar$.
The dynamics of the relative phase $\phi$ can induce oscillations in the phase difference 
$\psi\equiv (\theta_1-\theta_2)/2$. 
Let $\phi_R\equiv\phi_{1}^R-\phi_{2}^R$, $\phi_L\equiv\phi_{1}^L-\phi_{2}^L$, Fig.~\ref{fig1}~(c),
so that $\psi=(\phi_R-\phi_L)=\psi_0+\tilde\psi(t)$, where $\psi_0$
is the equilibrium value of $\psi$, and  $\tilde\psi(t)$ the time dependent part.
We can write $\theta_1=\theta_A+\psi$, $\theta_2=\theta_A-\psi$,
with $\theta_A=(\theta_1+\theta_2)/2$. Notice that $d\theta_A/dt = 2eV/\hbar$.
We consider the case when $V(t)=V_{dc} + V_{ac}\cos{\omega t}$.
When the Leggett mode is driven, directly or indirectly, by a periodic drive, 
we can assume $\tilde\psi\approx \hat A_{\omega}\sin(\omega t)$,
with $\hat A_{\omega}\approx A_0\Gamma_L\omega/((\omega^2-\omega_L^2)^2+\Gamma_L^2\omega^2)$ 
the amplitude of the mode and $\Gamma_L$ its broadening.
With this assumption we obtain
$\theta_1(t) = \theta_{0}            + (2e/\hbar)V_{dc}t + (2e/\hbar)(V_{ac}/\omega)\sin(\omega t) + \psi_0 + \hat A_\omega \sin(\omega t)$,
$\theta_2(t) = \theta_{0} + \phi_{0} + (2e/\hbar)V_{dc}t + (2e/\hbar)(V_{ac}/\omega)\sin(\omega t) - \psi_0 - \hat A_\omega \sin(\omega t)$,
where $\theta_0$ is an initial phase.
Let's first consider the case then $\omega\neq\omega_L$ and so $\hat A_\omega\approx 0$.
In this case we obtain
\begin{align}
 I = \sum_{n=0}^\infty (-1)^n&\left[
 	     I_1 J_n(\alpha_1 (2e/\hbar)V_{ac}/\omega)\sin(\theta_{0} + \psi_0 + \alpha_1 (2e/\hbar)V_{dc}t - n\omega t) \right. + \nonumber \\
	     &\left. I_2 J_n(\alpha_2 (2e/\hbar)V_{ac}/\omega)\sin(\theta_{0} - \psi_0 + \alpha_2 (2e/\hbar)V_{dc}t - n\omega t) 
         \right] 
  \label{eq:I}
\end{align}
where $n$ is an integer and $J_n(x)$ is the n-th Bessel function of the first kind.
The d.c. current will be zero unless
$V_{dc}=(\hbar/2e)n\omega/\alpha_i$, in which case the current exhibits so called Shapiro spikes~\cite{shapiro_josephson_1963}.
For the experimentally more relevant case when the JJs is current biased 
the Shapiro spikes correspond to steps, Shapiro steps, for the d.c. voltage.
If either $\alpha_1$ or $\alpha_2$ is equal to one we have Shapiro spikes for all integer values $n$.
If both $\alpha_1$ and $\alpha_2$ are equal to 1/2 Eq.~\ceq{eq:I} shows that 
we have Shapiro spikes only when $n$ is an even integer.
In the non-ideal case, 
it is possible that also in this case spikes will be present for odd $n$~\cite{Dominguez2017}.
Equation~\ceq{eq:I} shows that in general, regardless of the values of $\alpha_i$,
and the phase difference $\phi_0$, in the limit when $\hat A_\omega\approx 0$ 
we cannot have a suppression of the even steps alone. The reason is that when $\hat A_\omega\approx0$
the driving a.c. term $V_{ac}/\omega)\sin(\omega t)$ for $\theta_1$
is in phase with the driving term for $\theta_2$.

In the limit when $\omega\approx\omega_L$  so that $\hat A_{\omega_L} \gg V_{ac}/\omega$ instead of Eq.~\ceq{eq:I} we obtain:
\begin{align}
 I = \sum_{n=0}^\infty (-1)^n&\left[
 	                    I_1 J_n(\alpha_1 \hat A_{\omega_L})\sin(\theta_{0} + \psi_0 + \alpha_1 (2e/\hbar)V_{dc}t - n\omega_L t) \right. + \nonumber \\
	     &\left. (-1)^n I_2 J_n(\alpha_2 \hat A_{\omega_L})\sin(\theta_{0} - \psi_0 + \alpha_2 (2e/\hbar)V_{dc}t - n\omega_L t) 
         \right] 
  \label{eq:I2}
\end{align}
When $\alpha_1=\alpha_2=1$,
depending on the value of $\psi_0$ we can have suppression of the odd or even spikes.
For $\psi_0=0$ we have suppression of the odd steps. 
In this case for $\omega\approx\omega_L$ the Shapiro steps' structure is qualitatively the same as the one obtained
at low frequencies and powers in the presence of
a topological superconducting channel ($\alpha_1=\alpha_2=1/2$), or Landau-Zener processes
in highly transparent junctions~\cite{Dartiailh2021}. 
For small $\omega_L$ and non-negligible $\Gamma_L$ it might be difficult to pinpoint reliably 
the cause of the missing odd-Shapiro steps.
However, for the case when  $\psi_0=\pi/2$ we have that Eq.~\ceq{eq:I2} leads to a suppression of the even Shapiro spikes, a
phenomenon that cannot be attributed to the topological nature of the JJ, or Landau-Zener processes.
A phase difference $\psi_0\neq 0$ can be present, for example, due to strong spin-orbit coupling that causes 
the spin textures on the two bands to be different.
In the remainder we assume $\alpha_1=\alpha_2=1$.

To describe the dynamics of an a.c. current-biased 2-bands 
JJs we use a resistively and capacitively shunted junction (RCSJ) model~\cite{Barone1982}.
We assume that the external leads couple strongly to the states in band 1 and weakly to
the states in band 2. 
This situation is realized, for instance, when placing a lead on the surface of a DSM:
the states of the lead couple strongly to the DSM's surface states and weakly to the DSM's bulk states.
In this scenario the supercurrent in the bulk band (band 2) is mediated by intraband processes. 
The capacitance between the two leads is very small compared to the normal resistances $R_i$ across the leads and so it can can be neglected.
Conversely, for the inter band charge flow, within the same lead, 
we can neglect the resistive channel considering the non-negligible 
inter band capacitance $C_{12}$. 
The resulting effective RCSJ model is shown in Fig.~\ref{fig1}~(d).

In presence of the current bias $I_B=I_{dc}+I_{ac}\cos(\omega t)$, 
the dynamics of the RCSJ model shown in Fig.~\ref{fig1}~(c) is described by the equations~\cite{SI}
 \begin{align}
  &\frac{d\theta_A}{d\tau} = \xi \frac{d\psi}{d\tau} + i_{B}(\tau) - \sin \theta_1 - i_2 \sin \theta_2
  \label{eq:JJ1} \\
  &\frac{d^2\tilde\psi}{d\tau^2} + \frac{\omega_L^2}{\omega_J^2} \tilde\psi \approx \hat A_0 i_{ac} \cos(\hat\omega \tau)
  \label{eq:JJ2}
\end{align}
where $\omega_J\equiv 2eR I_1/\hbar$,
$\tau \equiv \omega_J t$
$R=R_1R_2/(R_1+R_2)$,
$\xi\equiv (R_1-R_2)/(R_1+R_2)$,
$\hat\omega\equiv\omega/\omega_J$
$i_B\equiv I_B/I_1$
$i_2\equiv I_2/I_1$,
and 
$\hat A_0\equiv \omega_L^2 R_1/(\omega_J^2 i_{12} (R_1+R_2))$.
In the remainder we set $\omega_L/\omega_J=0.005$, $\Gamma_L/\omega_L=7.5\cdot 10^{-5}$, $\hat A_0=0.0045$, $\xi=-0.6$, $i_2=1.5, 
$and $\beta=0.05\pi$.

The dynamics of the SQUID can be obtained starting from Eqs.~\ceq{eq:JJ1},~\ceq{eq:JJ2} for each of the two JJs.
In the remainder we will denote by
$X_i^{(j)}$ the quantity $X$ for band $i$ in arm $j$ of the SQUID, see Fig.~\ref{fig1}~(c).
We assume the SQUID to be symmetric: the parameters entering the JJs' RCSJ model, and the self inductance $L$,
are assumed to be the same for the left and right arm of the SQUID.
For each band the phase difference $\eta\equiv(\theta_i^{(2)}-\theta_i^{(1)})/2\pi$ must be
equal to $\Phi_{ext}/\Phi_0+m +\beta(i^{(1)}-i^{(2)})$,
where 
$\Phi_{ext}$ is the external flux threading the SQUID, 
$\Phi_0=h/2e$,
$m$ is an integer,
$\beta = I_1 L/\Phi_0$, 
and $i^{(j)}=I^{(j)}/I_1$ with $I^{(j)}$ the total current flowing through arm $j$.
In the remainder we set $m=0$.
In the limit in which $L$ is small, so that $\beta\ll 1$, 
we can assume 
$\eta = \hat\Phi + \beta \tilde\eta + \mathcal{O}(\beta^2)$~\cite{Romeo2005}, with $\hat\Phi=\Phi_{ext}/\Phi_0$.
Using Eqs.~\ceq{eq:JJ1},~\ceq{eq:JJ2}, current conservation, and the flux quantization for $\eta$, in the limit $\beta\ll 1$,
in terms of the phases 
$\theta_S\equiv\sum_{ij}\theta_i^{(j)}/4$,
$\psi=\psi^{(1)}=\psi^{(2)}=\psi_0+\tilde\psi(t)$, 
we find~\cite{SI} that the dynamics of the SQUID is described by the equations
\begin{align}
    &\frac{d\theta_S}{d\tau} = \xi \frac{d\psi}{d\tau} + \frac{1}{2}[i_B - i_s\left( \theta_s,~\psi \right)] 
      \label{eq:squid1} \\
    & i_s(\theta_s,~\psi)  = 2\cos(\pi\hat{\Phi}) 
                             \left[ \sin\left(\theta_s + \psi \right) + i_{2} \sin(\theta_s - \psi) \right]    
                            -2\beta \sin^2( \pi\hat{\Phi})
         \left[\sin(2(\theta_s + \psi)) + i_{2}^2 \sin(2(\theta_s - \psi)) + 2i_{2} \sin( 2\theta_s) \right]
         \label{eq:squid2}
\end{align}
in conjunction with Eq.~\ceq{eq:JJ2}.
Equation~\ceq{eq:squid2} shows that the SQUID's dynamics is $2\Phi_0$-periodic with respect to $\Phi_{ext}$.

\begin{figure}[ht!]
\centering
\includegraphics[width=0.98\columnwidth]{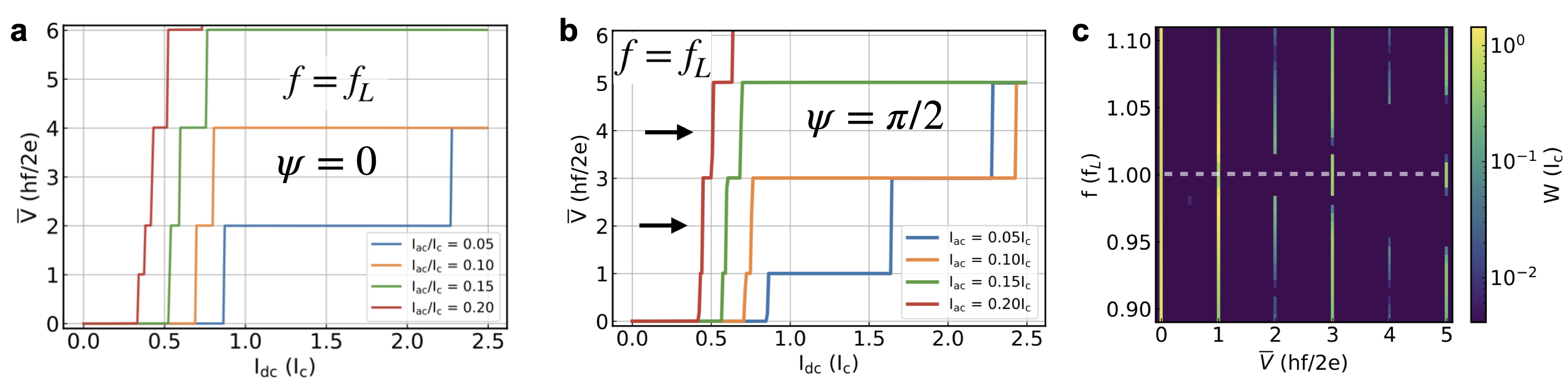}
\caption{\label{fig2}\textbf{Shapiro steps for a SQUID in presence of a Leggett mode when $\Phi_{ext}$=0.} 
 V-I curves for the case when $f = f_{L}$ and $\psi_0=0$, \textbf{a}, and $\psi_0=\pi/2$, \textbf{a}. 
 \textbf{c} Histogram of Shapiro steps as a function of ac frequency $f$ with power $I_{ac} = 0.05 I_1$.
}
\end{figure}

Using Eqs.~\ceq{eq:squid1},~\ceq{eq:squid2},~\ceq{eq:JJ2} we obtain $V(t)=(\hbar/2e)d\theta_s/dt$ and then 
$V_{dc}=\bar V =\lim_{t_f\to\infty}(1/t_f)\int_0^{t_f}V(t)dt$. 
Let's first consider $\hat\Phi=0 \mod 2$.
For $|\omega-\omega_L|\gg 1$,  the dependence of $V_{dc}$ with respect to $I_{dc}$ exhibits the standard
Shapiro steps~\cite{SI}: all steps if either $\alpha_1$, or $\alpha_2$ are equal to 1, only even steps if $\alpha_1=\alpha_2=1/2$,~\cite{SI}.
For $\omega=\omega_L$, $\psi_0=0$, and $\alpha_1=\alpha_2$
we have that the odd steps are strongly suppressed, see Fig.~\ref{fig2}~(a),
so that the structure of the Shapiro steps resembles the one expected
for a topological JJ for which a channel with $\alpha=1/2$ dominates.
However, for $\psi_0=\pi/2$, and $\alpha_1=\alpha_2=1$, we have the unusual situation that only the even Shapiro steps are suppressed,
as shown in Fig.~\ref{fig2}~(b).
This behavior is present as long as $\omega=2\pi f$ is within the inverse lifetime, $\Gamma_L$, 
of the Leggett mode frequency $f_L=\omega_L/2\pi$.
When $\hbar\omega_L=\hbar 2\pi f_L<\Delta_{sc}$ we can expect $\Gamma_L$ to be quite small.
Figure~\ref{fig2}~(c) shows the width of the steps, $W$, as a function of $V_{dc}$ and 
ac frequency $f$ assuming $\Gamma_L=0.05 f_L$. 
We see that for $|f-f_L|\ll \Gamma_L$
the even steps are suppressed while the odd steps
are strong, and that for $f$ far from the resonance we recover a voltage-current profile in which all the steps are present (apart from
small corrections due to higher harmonics).
 
We can investigate the effect of the Leggett mode on the Shapiro steps when
the SQUID is threaded by a nonzero magnetic flux $\Phi_{ext}$. 
For the case when $\hat\Phi\neq 0 \mod 2$, we first note
that for $\hat\Phi=1 \mod 2$
the second term vanishes.
In this case we find that the SQUID  V-I curve exhibits the same Shapiro steps as for the case $\hat\Phi= 0$.
When $\Phi_{ext}$ is a half-integer of $\Phi_0$ the first term on the r.h.s. of Eq.~\ceq{eq:squid2}
vanishes and the term proportional to $\beta$ affects the dynamics of the SQUID.
In this case, when $\psi\approx 0$, the factor of 2 in the argument of the sine causes the appearance
of half-integer Shapiro steps, as in standard SQUIDs~\cite{Vanneste1988}, when $\alpha_1=\alpha_2=1$, and the appearance of the odd Shapiro steps
when $\alpha_1=\alpha_2=1/2$.

\begin{figure}[ht!]
\centering
\includegraphics[width=0.98\columnwidth]{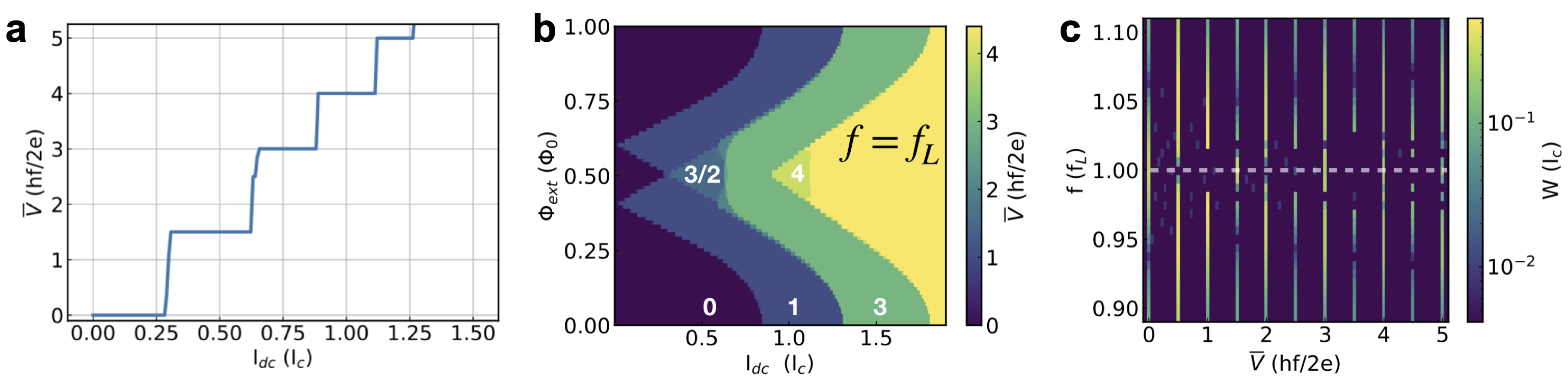}
\caption{\label{fig3}\textbf{Shapiro steps for a SQUID in presence of a Leggett mode when $\Phi_{ext}\neq 0$} 
 \textbf{a} Shapiro steps for a SQUID when $f=f_L$ and $\Phi_{ext}=\Phi_0/2$.
 \textbf{b} Colormap of Shapiro steps as a function of $\Phi_{ext}$. The different steps are labeled in white. 
 \textbf{c} Histogram of Shapiro steps as a function of ac frequency $f$. 
} 
\end{figure}

When $f\approx f_L$ so that $\tilde\psi$ is not negligible, and $\Phi_{ext}$ is not a multiple of $\Phi_0$, 
the SQUID's V-I features are difficult to predict from a simple analysis of the equations.
Numerically, for the case when $f=f_L$, $\psi_0=\pi/2$, and $\Phi_{ext}=\Phi_0/2$ we find that
the SQUID has a fairly unique V-I curve, as shown in Fig.~\ref{fig3}.
Contrary to the case of a single JJ the odd step at $V=(hf/2e)$ is absent, and
a new fractional step at $V=3/2 (hf/2e)$ appears together with a step at $V=4(hf/2e)$,
while the step at $V=3(hf/2e)$ survives. Figure~\ref{fig3}~(b) shows the range of values 
of $\Phi_{ext}$ around $\Phi_0/2$ for which these step structure is present,
and Fig.~\ref{fig3}~(c) shows how the step structure and the width of the steps depend on the a.c. frequency $f$, 
for $f\approx f_L$, when  $\Phi_{ext}=\Phi_0/2$. 

The discussion above shows that when the equilibrium phase difference, $\psi_0$, between the two superconducting
order parameters is 0 mod $2\pi$, the microwave response of a SQUID 
in which an undamped Leggett mode is present, for $\omega\approx\omega_L$ is similar to one obtained
when the single JJs forming the SQUID have a current phase relation that is $4\pi$ periodic, either
due to the presence of a topological superconducting channel, or Landau-Zener processes.
At the same, the analysis shows that when $\psi_0\approx\pi/2$ the SQUID's microwave response, both in
the absence and presence of an external magnetic flux $\Phi_{ext}$, 
exhibit unique qualitative features that cannot be attributed to topological superconducting
pairing or Landau-Zener processes.

\begin{figure}[ht!]
\centering
\includegraphics[width=0.9\columnwidth]{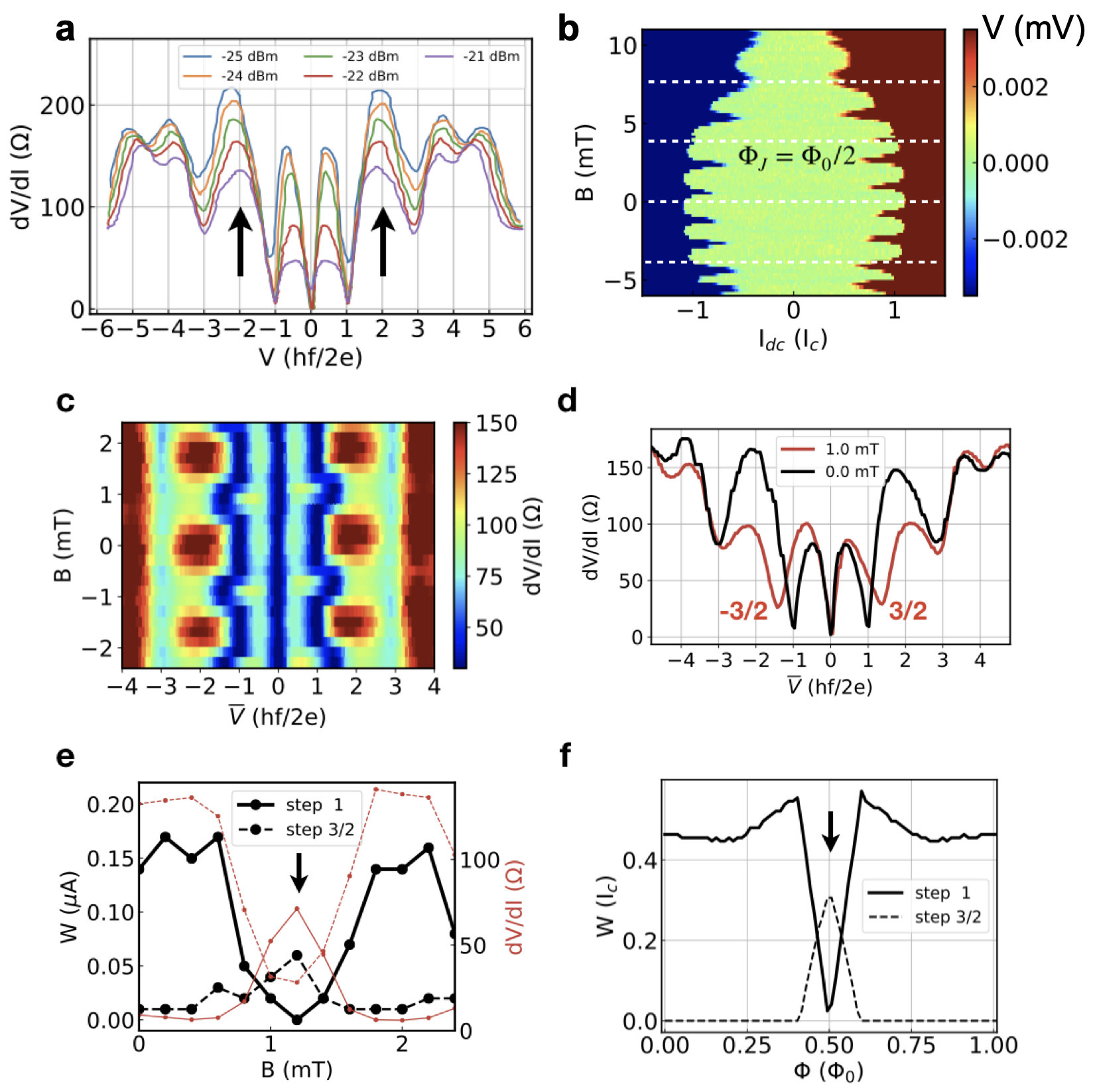}
\caption{\label{fig4}\textbf{Shapiro steps for a SQUID formed by a superconducting Dirac semimetal.}
 \textbf{a} Scans of the differential resistance for an a.c. driven SQUID's  at various powers, $f=9$~GHz, and $B=0$.
 \textbf{b} Anomalous SQUID oscillations in the d.c. regime occurring at integer multiples of $\Phi_0/2$, denoted by the dashed lines. 
 \textbf{c} Colormap of differential resistance versus $\bar V$ and $B$ for $f=9$ GHz and relative power -22 dBm. 
 \textbf{d} Comparison between Shapiro steps at zero field and $B = 1$ mT which corresponds to $\Phi_{ext} \sim \Phi_0/2$.
 \textbf{e} Measured Shapiro step widths, left vertical axis, and differential resistance, right vertical axis, vs $B$ for $f=9$ GHz.
 \textbf{f} Theoretical results for the Shapiro step widths vs flux threading the SQUID at $I_{ac} = 0.05I_c$ and $f=f_{L}$.
 }
\end{figure}

In a Dirac semimetal (DSM) the bulk 3D conduction and valence electronic bands touch at isolated points,
and a projection of the spectral density onto a surface BZ reveals Fermi arcs connecting the bulk Dirac points~\cite{Wehling2014,Armitage2018}. 
DSM's with proximity-induced superconductivity are predicted to be able to realize exotic non-Abelian anyons 
that can used to develop topologically-protected qubits \cite{Kitaev2001} 
and can be used in microwave single-photon detection for sensing applications \cite{Chi2020,Chatterjee2021,Pan2021}. 
Another aspect DSMs that has received less attention in the literature concerns the multiband properties of superconducting DSMs~\cite{Wang2018,Huang2019}. 
By placing a superconducting material on the surface of the Dirac semimetal  \cdas 
superconducting pairing can be induced in \cdas~\cite{Huang2019}.
The pairing has been shown to be characterized by two order parameters, $\Delta_1$, $\Delta_2$.
In addition, recent experiments on single JJs formed by superconducting
leads based on Al/\cdas have shown compelling signatures
that at equilibrium the phase difference, $\theta_1-\theta_2$, between the two phases across the junction
arising from the two superconducting order parameters is equal to $\pi$, implying $\psi_0=\pi/2$~\cite{Yu2018}.
Motivated by these results and the theoretical analysis above we have investigated the
microwave response of a SQUID based on Al/\cdas. 
Details about the fabrication and measurement of the device can be found in the Methods section and SI~\cite{SI}.

At low frequencies, $f<9$~GHz, and $\Phi_{ext}=0$, the SQUID's measured $dV/dI$ exhibits peaks and valleys consistent
with the standard Shapiro steps' structure (see SI~\cite{SI}). 
However, for $f=9$~GHz, for all the microwave powers values considered, 
the first and third steps are clearly visible but the second step is strongly suppressed, 
see Fig.~\ref{fig3}~(a). 
Considering that our device shows no hysteretic features in the current-voltage characteristic~\cite{SI},
and no evidence of a bias-dependent normal resistance,
mechanisms for missing Shapiro steps due to hysterisis~\cite{Shelly2020} or bias-dependent resistance~\cite{Mudi2021} are not relevant.
Based on the analysis above, we therefore conclude that
the results shown in  Fig.~\ref{fig3}~(a) can only be explained by considering the 
presence of a Leggett mode with $f_L\approx 9$~GHz, and $\psi_0=\pi/2$. 

Figure~\ref{fig4}~(b) shows the voltage across the SQUID as a function of the perpendicular magnetic field $B$ in the d.c. limit, $I_{ac}=0$.
SQUID oscillations of periodicity $\sim$1.8 mT are observed, which correspond to an effective SQUID ring area of $\sim$1.14 $\mu$m$^2$.
Enveloping the SQUID oscillations is the Fraunhofer diffraction pattern of the JJs. 
For $B$ such that $\Phi_{ext}=\Phi_0/2$ anomalous oscillations can also be observed.
The presence of these oscillations is consistent with a $\pi$-periodic supercurrent in each of the JJs forming the SQUID due
the fact that $\psi_0=\pi/2$. 

In Fig.~\ref{fig4}~(c) we present as a color plot the measured $dV/dI$ as a function of $\bar V$ and $B$
in the presence of an a.c. component of the current with $f=9$~GHz, and relative power -22~dB.
In addition, to the periodicity of the Shapiro steps with respect to B, with period consistent
with the periodicity observed in the d.c. limit, Fig.~\ref{fig4}~(b), we observe interesting features 
for $B\approx 1$~mT corresponding to $\Phi_{ext}=\Phi_0/2$. To more clearly identify these features,
in Fig.~\ref{fig4}~(d) we show the $dV/dI$ traces for $B=0$ and $B=1$~mT.
We see that for $B=1$~mT, i.e., $\Phi_{ext}=\Phi_0/2$,
both the first and second Shapiro steps are suppressed and
a 3/2 subharmonic step emerge, features that are remarkably consistent with the theoretical results shown in Fig.~\ref{fig3}. 
To better understand the evolution of the Shapiro steps' structure with $\Phi_{ext}$ when $f=9$~GHz, 
in Fig.~\ref{fig4}~(e) we plot the measured width of the steps at $V=hf/2e$ and $V=(3/2)(hf/2e)$
as a function of $B$. We see that when $B\approx 1$~mT, $\Phi_{ext}=\Phi_0/2$, the width of the first step is suppressed
 whereas the width of the 3/2 step is enhanced
around $B\approx 1$~mT. The evolution of the 1 and 3/2 steps with $\Phi_{ext}$ is 
in good qualitative agreement with the theoretical results, shown in Fig.~\ref{fig4}~(f). 

In conclusion, we have presented a theoretical analysis of the response of a SQUID to
microwave radiation in the presence of a Leggett mode, using an effective 
resistively and capacitively shunted junction model for Josephson junctions formed by superconducting leads with two order parameters. 
We have shown that, when the a.c. current's frequency $f$ is close to the frequency $f_L$
of the Leggett mode, the dc current-voltage characteristics of both a single junction and a SQUID 
exhibit suppressed odd Shapiro steps if the
equilibrium relative phase, $\psi_0$,
between the phases of the two superconducting order parameters is zero,
and strongly suppressed even Shapiro steps if $\psi_0=\pi/2$.
In the absence of hysteretic effects and spurious cavity resonances, the suppression of only the even-steps cannot be explained using other
mechanisms, such as the presence of channels for the JJs with a $4\pi$-periodic current phase relation.
In addition, we have shown that for $f\approx f_L$, and $\psi_0=\pi/2$, the evolution
of a SQUID's Shapiro steps structure with the magnetic flux $\Phi_{ext}$
possess unique qualitative features.
We have then presented experimental measurements taken on a SQUID formed by 
Josephson junctions whose superconducting leads are formed by Al placed on top 
of the Dirac semimetal \cdas. The results presented are consistent with Al inducing
a two-band superconducting  pairing in \cdas, with a $\psi_0=\pi/2$ difference
between the phases of the two superconducting order parameters, and 
the presence of a Leggett mode with $f_L\approx 9$~GHz.

The theoretical and experimental results presented show that 
the SQUIDs' response to microwave radiation
can be used to identify unambigously the presence of Leggett modes,
especially when the equilibrium phase difference between the phases
of the superconducting order parameters is not zero. In addition,
the results strongly suggest that in a Dirac semimetal like \cdas the multi-band superconducting
state induced via proximity effect exhibits an undamped Leggett mode.
In this study, we used Cd3As2 as an exemplar Dirac Semimetal,
but our results indicate that underdamped Leggett modes may be present also in other superconducting Dirac semimetals.


\vspace{0.5cm}

\noindent
\textbf{Methods}
\smallskip

\noindent
\textbf{Fabrication.} 
Mechanical exfoliation is used to obtain flat and shiny Cd$_3$As$_2$ thin flakes of thickness $\sim$200 nm from an initial bulk ingot material \cite{Pan2021}, synthesized via a chemical vapor deposition method\cite{Ali2014}. The SQUID structure is fabricated by first depositing the Cd$_3$As$_2$ thin flake on a Si/SiO$_2$ substrate with a 1 $\mu$m thick SiO$_2$ layer. Next, e-beam lithography is used to define 300 nm thick Al electrodes. Additional details about the device can be found elsewhere \cite{Yu2018}. 

\noindent
\textbf{Measurements.} 
To measure the sample resistance, a $\sim$11 Hz phase-sensistive lock-in amplifier technique is used with an excitation current of 10 nA. To measure the differential resistance, a large direct current up to $\pm$2 $\mu$A is added to the ac current. The entire device is emmersed in a cryogenic liquid at a temperature of $\sim$0.25 K, well below the devices superconducting transition temperature. To measure the microwave response of the device, an Agilent 83592B sweep generator is used to generate microwaves, which are conducted through a semirigid coax cable. 

\noindent
\textbf{Simulations.}
The numerical integration of the dynamical equations have been performed using the adaptive Runge-Kutta of order 4 and 5.

\noindent
\textbf{Data availability}

\noindent
The data that support the findings of this study are available within the paper and its Supplementary Information.

All the codes used to obtain the numerical results presented are available upon request.\\




%


\vspace{\baselineskip}

\noindent
\textbf{Acknowledgements}

\noindent
J.J.C, W.P., and E.R. acknowledge support from DOE, Grant No DE-SC0022245. J.J.C. also acknowledges support from the Graduate Research Fellowship awarded by the Virginia Space Grant Consortium (VSGC). The authors acknowledge the \cdas material synthesis work of David X. Rademacher. The work at Sandia is supported by a LDRD project. Device fabrication was performed at the Center for Integrated Nanotechnologies, a U.S. DOE, Office of BES, user facility. Sandia National Laboratories is a multimission laboratory managed and operated by National Technology and Engineering Solutions of Sandia LLC, a wholly owned subsidiary of Honeywell International Inc.~for the U.S.~DOE's National Nuclear Security Administration under contract DE-NA0003525. This paper describes objective technical results and analysis. Any subjective views or opinions that might be expressed in the paper do not necessarily represent the views of the U.S. DOE or the United States Government. \\

\noindent
\textbf{Authors contributions}

\noindent
J.J.C. and E.R. developed the theoretical model. J.J.C. carried out the numerical simulations. 
W.Y., P.D., T.M.N., D.B.S., and W.P. conceived the experiment, and contributed to material growth, device fabrication, electronic transport measurements, and experimental data analysis.  W.P. coordinated the experiment. 
All authors contributed to interpreting the data. 
The manuscript was written by J.J.C., W.P., and E.R. with suggestions from all other authors.


\newpage

\setcounter{section}{0}

\renewcommand{\thefigure}{S\arabic{figure}}
\setcounter{figure}{0}    

\renewcommand{\theequation}{S\arabic{equation}}
\setcounter{equation}{0}  

\begin{center}
\large{\bf SUPPLEMENTAL MATERIAL}
\end{center}


\section{Dynamics of a two-bands Josephson junction in the presence of a Leggett mode}

Let's consider a Josephson junction (JJ) where each of the superconducting electrodes are two-band superconductors with phases $\phi_1,~ \phi_2$.
Let the intraband phase differences across the junction be $\theta_i = \phi_i^{R} - \phi_i^L$. 
To describe the \textit{interband} dynamics in the JJ, we consider the resistively and capacitively shunted junction (RCSJ) 
model shown in Fig.~\ref{fig1}~(d). 
For the ac Josephson effect we have that the voltage $V$ across a weak link, denoted by crosses in Fig.~\ref{fig1}~(d), is given by 
$V = \hbar \dot{\varphi}_i/2e$, where $\varphi$ is the phase difference across the weak link of the superconducting order parameters.
Let $\phi^L = \phi_1^L - \phi_2^L$,  $\phi^R = \phi_1^R - \psi_2^R$ and $\theta_i = \phi_i^R-\phi_i^L$.
From Kirchhoff's voltage law applied to the loop formed by the weak links in Fig.~\ref{fig1}~(d) we obtain
\beq
 \dot{\phi}_L + \dot{\theta}_2 - \dot{\phi}_R - \dot{\theta}_1 = 0
 \label{eq:k01}
\enq 
and then $\dot{\theta}_2 - \dot{\theta}_1= \dot{\phi}_R - \dot{\phi}_L$.

Let $I_{12}^L$, $I_{12}^R$ be the interband critical dc Josephson current on the left side and right side,
respectively, of the circuit shown in Fig.~\ref{fig1}~(d) of the main text,
and $C_{12}^L$, $C_{12}^R$ the left-side, right-side, interband capacitances.
From charge conservation we obtain
\begin{align}
            & I_{12}^L \sin\phi^L + \frac{\hbar}{2e}C_{12}^L \ddot{\phi}^L = I_2 \sin \theta_2 + \frac{\hbar}{2e R_2}\dot{\theta}_2    \label{eq:psiL} \\
            & I_{12}^R \sin\phi^R + \frac{\hbar}{2e}C_{12}^R \ddot{\phi}^R = -(I_2 \sin \theta_2 + \frac{\hbar}{2e R_2}\dot{\theta}_2) \label{eq:psiR}
\end{align}
If we assume $C_{12}^L = C_{12}^R \equiv C_{12}$ and $I_{12}^L = I_{12}^R \equiv I_{12}$, 
it is clear that $\phi^R = - \phi_L$. Then, from Eq.~\ceq{eq:k01} we obtain
\beq
 \psi \equiv \frac{\theta_1-\theta_2}{2} = \phi_L = -\phi_R.
 \label{eq:psi01}
\enq 
Equation~\ceq{eq:psi01} establishes the direct relation between phases across the Josephson junction, $\theta_i$, and
the phases $\phi_R$, $\phi_L$, characterizing the Leggett modes in the two superconducting leads. 
In particular Eq.~\ceq{eq:psi01} implies that the dynamics of the Leggett modes will in general affect
the dynamics of the phases across the JJ. 

We now obtain the dynamics of the current biased JJ shown in Fig.~\ref{fig1}~(d) taking into
account the presence of a Leggett mode. 
When a bias current $I_B$ is applied across the junction, charge conservation gives
\begin{align}
  I_B &= I_1 \sin \theta_1 + I_2 \sin \theta_2 + \frac{V_1}{R_1} + \frac{V_2}{R_2}
\label{eq:cons_charge}
\end{align}
where $I_i$ is the critical Josephson current for the $i^{th}$ band and $V_i/R_i$ is the current through the resistive channel in the $i^{th}$ band. 
Let $\theta_A \equiv (\theta_1+\theta_2)/2$. Considering Eq.~\ceq{eq:psi01}, we can write $\theta_1=\theta_A+\psi$, $\theta_2=\theta_A-\psi$ and then
\begin{align}
  I_B &= I_1 \sin \theta_1 + I_2 \sin \theta_2 + \frac{\hbar}{2eR} \left( \dot{\theta}_A - \xi \dot{\psi} \right)
  \label{eq:IB}
\end{align}
where 
$R = R_1 R_2/(R_1 + R_2)$ is the parallel resistance of the resistors $R_1$ and $R_2$, 
$\xi = (R_1 - R_2)/(R_1 + R_2)$ quantifies the asymmetry in resistance between the bands. 
Defining $\omega_J\equiv 2eR I_1/\hbar$,
$\tau \equiv \omega_J t$, we can write Eq.~\ceq{eq:IB} as
\begin{align}
  \frac{d\theta_A}{d\tau} = \xi \frac{d\psi}{d\tau} + i_B - \sin \theta_1 - i_2 \sin \theta_2
  \label{eq:thetaA}
\end{align}
where currents have been normalized with respect to $I_1$: $i_B = I_B/I_1$ and $i_2 = I_2/I_1$.
Equation~\ceq{eq:thetaA} is the key equation to describe the behavior of the 2-band JJ, and SQUID (see next section), in the
presence of a Leggett mode. The key modification due to the Leggett mode is
the term $\xi d\psi/d\tau$. The evolution in time of $\psi(t)$ depends on several microscopic details
that are beyond the level of the effective description used here. 
We have assumed $\psi(t)$ to follow the dynamics of a harmonic oscillator driven by a periodic term due to the microwave radiation.
Below we show that, in first approximation, this simplified evolution is also consistent with the RSCJ model 
shown in Fig.~\ref{fig1}~(d).

We can rewrite Eq.~\ceq{eq:psiR} in the form:
\begin{equation}
 \frac{d^2\psi}{d\tau^2} + \frac{R\omega_L^2}{\omega_J^2 i_{12}(R_1+R_2)}\frac{d\psi}{d\tau} + \frac{\omega_L^2}{\omega_J^2}\sin\psi = 
 \frac{\omega_L^2 R_1}{\omega_J^2 i_{i2} (R_1+R_2)}\left[i_B + (R_2/R_1) i_2\sin(\theta_A -\psi) - \sin(\theta_A+\psi) \right].
 \label{eq:psi}
\end{equation}
where $i_{12}\equiv I_{12}/I_1$, $\omega_L=\sqrt{(2e/\hbar) I_{12}/C_{12}}$ is the Leggett mode's frequency, 
and $i_B=i_{dc}+i_{ac}\cos(\omega\tau)$
Equations~\ceq{eq:thetaA} and~\ceq{eq:psi} completely define the dynamics of the two-bands JJ described by the effective RCJS circuit
shown in Fig.~\ref{fig1}~(d).
Eq.~\ceq{eq:psi} is equivalent to the equation for a damped, driven, oscillator; the right hand side of the equation
being the driving term. To qualitatively understand the effect of a resonant Leggett mode,
in first approximation, we can neglect the damping term proportional $d\psi/d\tau$, and the term 
$i_{dc} + (R_2/R_1) i_2\sin(\theta_A -\psi)-\sin(\theta_A+\psi)$ 
on the right hand side of the equation. Then by linearizing the $\sin\psi$ around the equilibrium value $\psi_0$
for $\tilde\psi\equiv \psi-\psi_0$ we obtain the simple equation
\begin{equation}
 \frac{d^2\tilde\psi}{d\tau^2}  + \frac{\omega_L^2}{\omega_J^2}\tilde\psi = 
 \frac{\omega_L^2 R_1}{\omega_J^2 i_{i2}(R_1+R_2)}i_{ac}\cos(\hat\omega\tau)
 \label{eq:psi2}
\end{equation}
describing a harmonic oscillator periodically driven by a force of amplitude $\hat A_0 i_{ac}$, with
$\hat A_0\equiv \omega_L^2 R_1/(\omega_J^2 i_{12} (R_1+R_2))$. Here $\hat\omega\equiv\omega/\omega_J$.
In our calculations the effect of the damping term is taken into account by considering a finite broadening, $\Gamma_L$,
of the Leggett mode's resonance frequency.

    
\noindent
\section{Dynamics of a SQUID formed by two-bands superconducting leads and in the presence of a Leggett mode.}
In this section we derive the equations that we use to simulate the dynamics of a two-bands SQUID
in presence of a resonant Leggett mode.
We assume the SQUID to be symmetric:
\begin{align}
    C_{12}^{(1)} & = C_{12}^{(2)} \equiv C_{12}; \quad R_i^{(1)} = R_i^{(2)} \equiv R_i; \quad I_i^{(1)} = I_i^{(2)} \equiv J_i; \nonumber \\
    I_{12}^{(1)} & = I_{21}^{(1)} = I_{12}^{(2)} = I_{21}^{(2)} \equiv I_{12}, \nonumber
\end{align}
where $X_i^{(j)}$ denotes quantity $X$ in band $i$, and arm $(j)$ of the SQUID. Normalizing as usual the currents with $I_1$,
from charge conservation and magnetic flux quantization we have:
\begin{align}
  & i^{(1)} + i^{(2)} = i_B  \label{eq:isum} \\
  & i^{(1)} - i^{(2)} =\frac{\theta_i^{(2)}-\theta_i^{(1)}}{2\pi\beta} -\frac{\hat\Phi}{\beta} + \frac{m}{\beta} \label{eq:idiff}
\end{align}
where $\beta\equiv I_1L/\Phi_0$, $\hat\Phi\equiv\Phi_{ext}/\Phi_0$, and $m$ is an integer that without loss of generality we can set equal to zero.
For the total current in arm $(j)$ we have:
\beq
 i^{(j)} = \frac{\hbar}{2e R_1 I_1}\frac{d\theta_1^{(j)}}{dt} + \frac{\hbar}{2e R_2 I_1}\frac{d\theta_2^{(j)}}{dt} +
           \sin(\theta_1^{(j)}) + i_2^{(j)}\sin(\theta_2^{(j)}).
           \label{eq:ij}
\enq
Let's now define
\begin{equation}
 \psi^{(1)}\equiv \frac{\theta_1^{(1)}-\theta_2^{(1)}}{2}; \hspace{0.5cm} 
 \psi^{(2)}\equiv \frac{\theta_1^{(2)}-\theta_2^{(2)}}{2}; \hspace{0.5cm} 
 \eta_1\equiv \frac{\theta_1^{(2)}-\theta_1^{(1)}}{2\pi};  \hspace{0.5cm} 
 \eta_2\equiv \frac{\theta_2^{(2)}-\theta_2^{(1)}}{2\pi};  \hspace{0.5cm} 
 \theta_S =\frac{1}{4}\sum_{ij}\theta_i^{(j)}.
\end{equation}
Because the flux quantization condition is the same for both bands, we have $\eta_1=\eta_2\equiv\eta$, and $\psi^{(1)}=\psi^{(2)}\equiv\psi$.
$\psi$ is the phase associated to the Leggett mode and its dynamics is given by Eq.~\ceq{eq:psi2}.
By using Eq.~\ceq{eq:ij} to express $i^{(j)}$ in Eqs.~\ceq{eq:isum},~\ceq{eq:idiff} we obtain the following dynamical equations for $\theta_S$ 
and $\eta$
\begin{align}
 &\frac{d\theta_S}{d\tau} - \xi \frac{d\psi}{d\tau} = \frac{i_B}{2} - \frac{1}{2}i_s(\theta_s,\psi,\eta)
 \label{eq:thetas} \\
 &2\pi\frac{d\eta}{d\tau}     = -\frac{\eta}{\beta} + \frac{\hat\Phi}{\beta} + i_{d}(\theta_s,\psi,\eta)
 \label{eq:eta} 
\end{align}
where 
\begin{align}
 i_s(\theta_s,\psi,\eta)     & =       \sin\theta_1^{(1)} + \sin\theta_1^{(2)} + i_2 [\sin\theta_2^{(1)} + \sin\theta_2^{(2)}] \nonumber \\
                             & =       \sin(\theta_S + \psi - \pi\eta) + \sin(\theta_S + \psi + \pi\eta) +
                                       i_2[\sin(\theta_S - \psi - \pi\eta) + \sin(\theta_S - \psi + \pi\eta)];  
 \label{eq:is01}\\
 i_{d}(\theta_s,\psi,\eta) & =      \sin\theta_1^{(1)} - \sin\theta_1^{(2)} + i_2 [\sin\theta_2^{(1)} - \sin\theta_2^{(2)}] \nonumber \\
  	                         & =      \sin(\theta_S + \psi - \pi\eta) - \sin(\theta_S + \psi + \pi\eta) +
                              i_2[\sin(\theta_S - \psi - \pi\eta) - \sin(\theta_S - \psi + \pi\eta)]
                              \label{eq:is12} 
\end{align}
$i_s$ is the supercurrent fraction of the total current across the SQUID.
In the limit $\beta\ll 1$ we can assume~\cite{Romeo2005} 
\beq
 \eta = \hat\Phi + \beta\tilde\eta + \mathcal{O}(\beta^2).
 \label{eq:eta2}
\enq
From Eq.~\ceq{eq:eta}, for $\tilde\eta$, we find:
\beq
 \tilde\eta = 2\sin(\pi\hat\Phi)[\cos(\theta_S+\psi) + i_2 \cos(\theta_S-\psi)] + \mathcal{O}(\beta)]
 \label{eq:tildeeta}
\enq
Replacing in the equation~\ceq{eq:is01} for $i_s$
the expression for $\eta$ obtained by combining Eqs.~\ceq{eq:eta2},~\ceq{eq:tildeeta}, we obtain, to linear order in $\beta$:
\begin{align}
 i_s(\theta_s,\psi) = & 2\cos(\pi\hat\Phi)[\sin(\theta_S + \psi) + i_2\sin(\theta_S - \psi)] - \nonumber \\
                      & 2\beta\sin^2(\pi\hat\Phi)[\sin(2(\theta_S + \psi)) + i_2^2\sin(2(\theta_S - \psi)) + 2 i_2\sin(2\theta_S)]
 \label{eq:is02}.
\end{align}
Notice that up to linear order in $\beta$ $i_s$ only depends on $\theta_S$ and $\psi$.

Equations~\ceq{eq:thetas},~\ceq{eq:is02}, and~\ceq{eq:psi2} completely determine the dynamics of the SQUID.
To numerically integrate these non-linear differential equations we used an adaptive fourth-order Runge-Kutta method. 
The parameters of the model used in the simulations are given in Table~\ref{tab:params}.
\begin{table}[h!]
\centering
\begin{tabular}{||c | c | c | c | c | c ||} 
 \hline
 $\omega_L/\omega_J$ & $\Gamma_L/\omega_L$ & $\hat A$ & $\xi$  & $i_2$ & $\beta$ \\ [0.5ex] 
 \hline
 0.005 & $7.5\cdot 10^{-5}$ & $4.5\cdot 10^{-3}$ & -0.6 & 1.5 & $0.05\pi$ \\
 \hline 
\end{tabular}
\caption{\label{tab:params}}
\end{table}
%

\section{Additional theoretical results}
\begin{figure}[h!]
\centering
\includegraphics[width=0.80\textwidth]{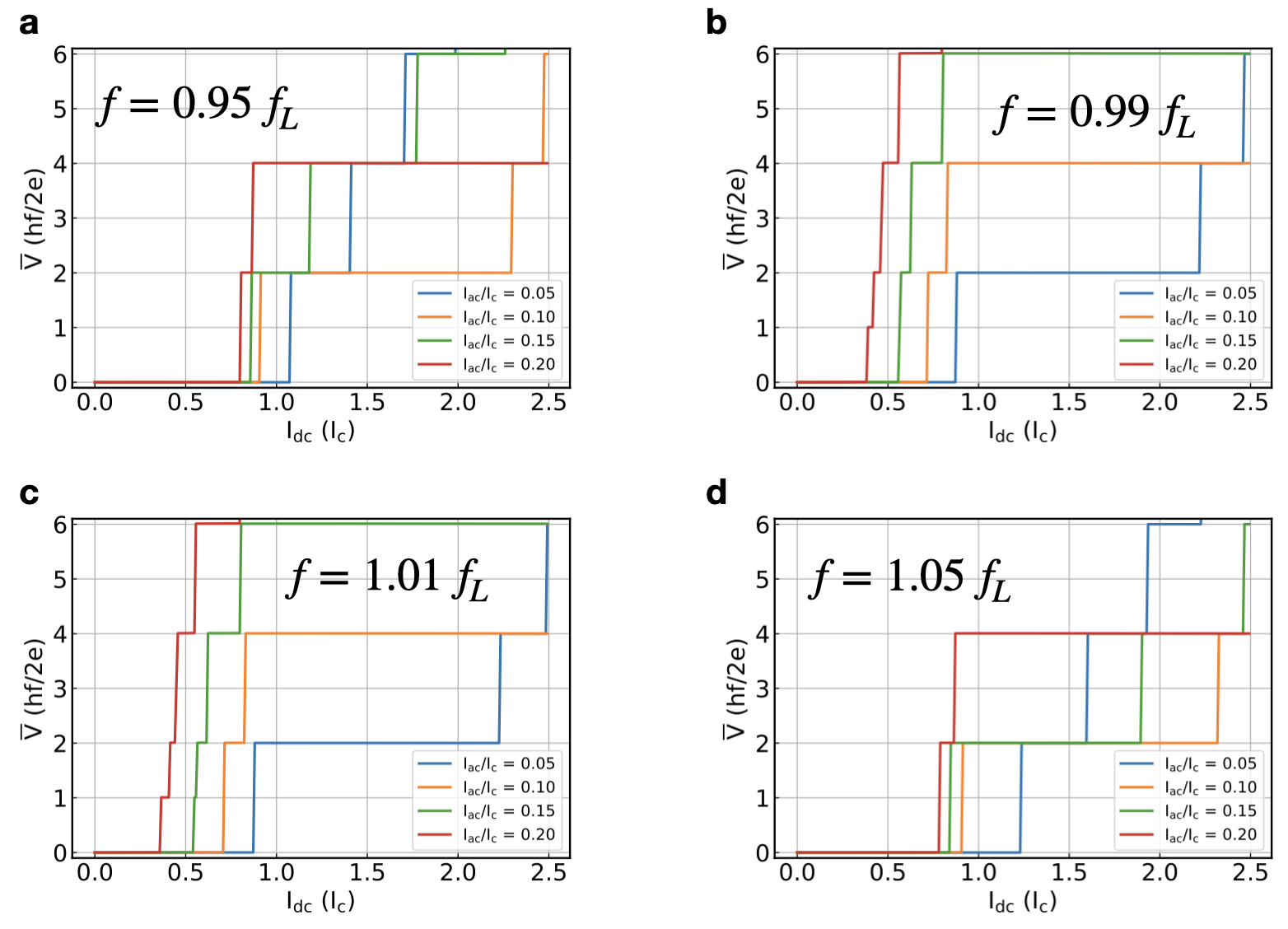}
\caption{\label{fig:missing_odd} Shapiro steps at various ac frequencies for $\Phi=0$ and using the same parameters as those used to generate other calculations except that here the intrinsic phase between the two bands in a given junction is zero (rather than $\pi$).}
\end{figure}

In Fig.~\ref{fig:missing_odd} we present additional numerical VI curves in the case where $\psi_0 = 0$. Here, we see, as mentioned in the main text, the missing steps are odd integer multiples of $(hf/2e)$. The ac frequency range in Fig.~\ref{fig:missing_odd}a-d is chosen to cover the approximate half-width of the Leggett mode resonance in the amplitude $A_{\omega} = A_0 \Gamma_L \omega / ((\omega^2 - \omega_L^2)^2 + \Gamma_L^2 \omega^2)$, illustrating the robustness of the missing steps over a bandwidth proportional to the inverse lifetime of the Leggett mode.

\begin{figure}[h!]
\centering
\includegraphics[width=0.80\textwidth]{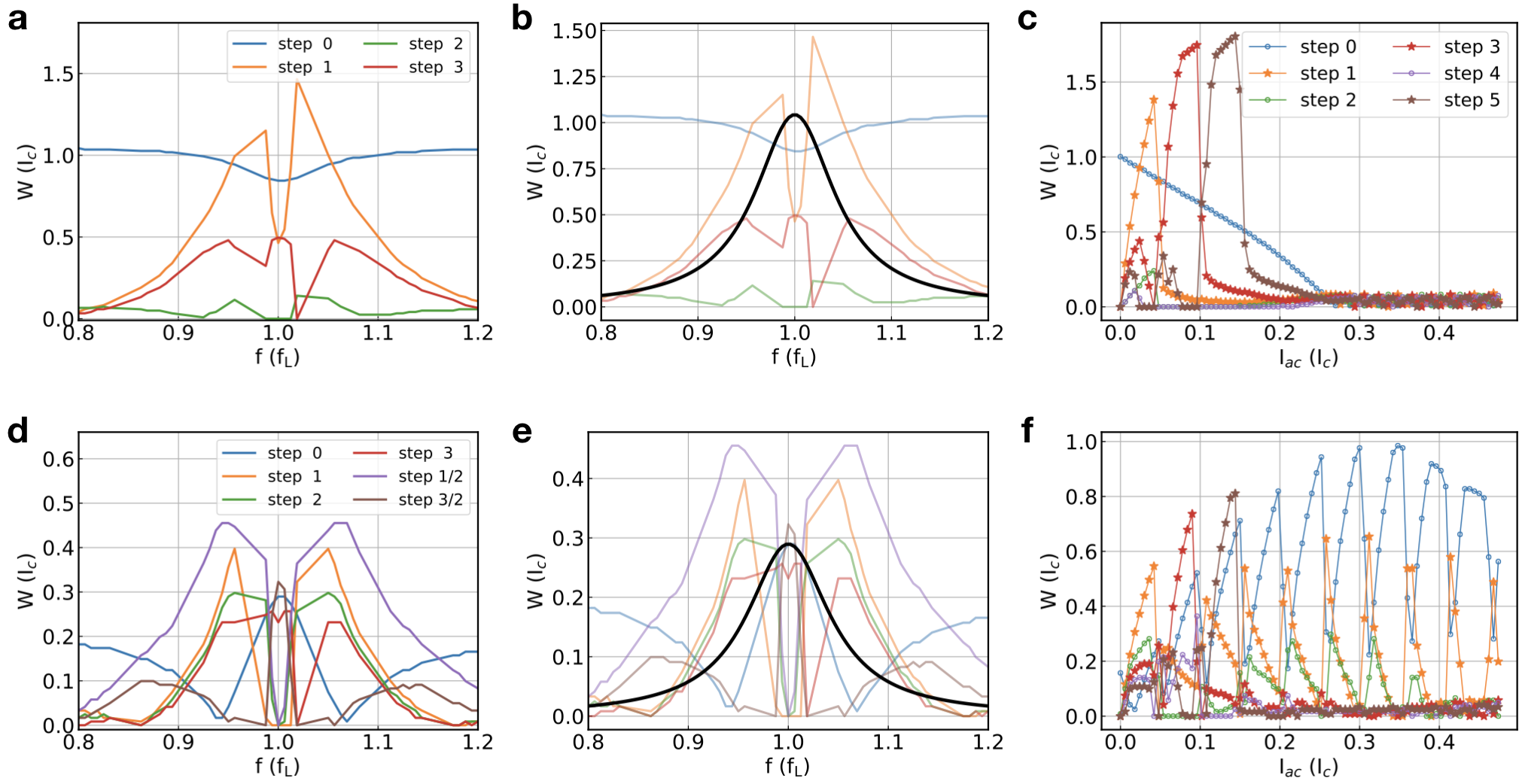}
\caption{\label{fig:freq_power} \textbf{a-b} Step width dependence on ac frequency for $\Phi =0$ and $I_{ac} = 0.05 I_c$. The bold black line represents the Lorenzian corresponding to the Leggett mode linewidth. \textbf{c} Step width dependence on ac power for $\Phi =0$ and $f = f_{Leggett}$. \textbf{d-e} Step widths dependence on ac frequency for $\Phi =\Phi_0/2$ and $I_{ac} = 0.05 I_c$. \textbf{f} Step width dependence on ac power for $\Phi =\Phi_0/2$ and $f = f_{Leggett}$.}
\end{figure}

In Fig.~\ref{fig:freq_power}, we present calculations of Shapiro step widths of the $n^{th}$ step corresponding to $\overline{V} = n(hf/2e)$ in the case where $\psi_0 = \pi/2$. Fig.~\ref{fig:freq_power}a-b show the step width ac frequency dependence near the Leggett mode frequency and for $\hat{\Phi} = 0$ and $I_{ac}/I_c = 0.05$, where a normalized $A_{\omega}$ is shown in black for reference. Clearly, deviations from the conventional Shapiro step dependence follows the resonant Leggett amplitude. Fig.~\ref{fig:freq_power}c show the power dependence of steps for $\hat{\Phi} = 0$ and $f = f_L$. We see the gap is suppressed at $I_{ac} \sim 0.25~I_c$, which is much smaller than expected in the conventional case. Furthermore, the step width dependence of odd steps exhibit resonant features appearing consecutively with increasing power and disappering with the gap closure. Once the gap is closed, step widths exhibit oscillations in power, similar to the conventional Bessel regime.

In Fig.~\ref{fig:freq_power}d-e, we show the step width ac frequency dependence near the Leggett mode frequency and for $\hat{\Phi} = 1/2$ and $I_{ac}/I_c = 0.05$, where a normalized $A_{\omega}$ is shown in black for reference. We observe a weakening of the gap near the Leggett frequency, similar to the zero-flux case, but the gap actually becomes enhanced at the Leggett frequency. In Fig.~\ref{fig:freq_power}f, we present the power dependence of steps for $\hat{\Phi} = 1/2$ and $f = f_L$. We find similar resonant behavior of odd steps at low power, but the features are difficult to distinguish between oscillations at higher powers associated with the typical Bessel oscillations.

\clearpage

\section{Device characterization}
\begin{figure}[h!]
\centering
\includegraphics[width=0.98\textwidth]{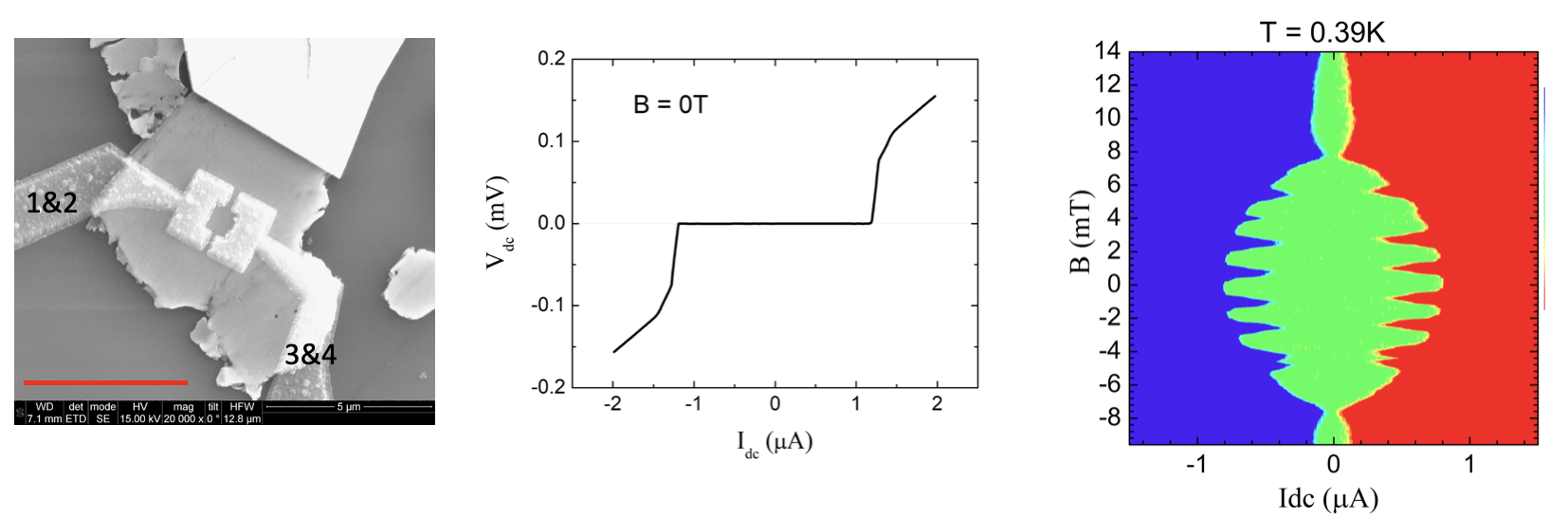}
\caption{\label{fig:device} \textbf{a} An SEM image of the SQUID device used in the experiment.
\textbf{b} The I-V curve measured at B = 0T. 
\textbf{c} The I-V curves as a function of out-of-plane magnetic fields, at a higher temperature of T = 0.39K.}
\end{figure}

Fig.~\ref{fig:device}a  shows an SEM image of the SQUID device used in the experiment. The scale bar is 5 $\mu$m. For each Josephson junction in the SQUID, the width is ~ 600 nm, and the gap ~ 150 nm. The size of the middle open square is about 800nm x 800nm. For IV and differential resistance measurements, the ac/dc current runs from contact 1 to contact 3. The dc/ac voltage is measured between contacts 2 and 4. In Fig.~\ref{fig:device}b, we present the I-V curve measured at B = 0T. The critical current is $\sim 1.1~\mu A$.
In Fig.~\ref{fig:device}c, we show I-V curves as a function of out-of-plane magnetic fields, at a higher temperature of T = 0.39K (compared to Fig.~4a in the main text). In this plot, red color represents positive Vdc, blue negative Vdc. In the green color regime, Vdc = 0. A typical feature, i.e., the envelop of the SQUID oscillatory pattern being modulated by the Fraunhofer diffraction pattern of the single Josephson junction, is clearly seen.

\section{Additional experimental results}
\begin{figure}[h!]
\centering
\includegraphics[width=0.98\textwidth]{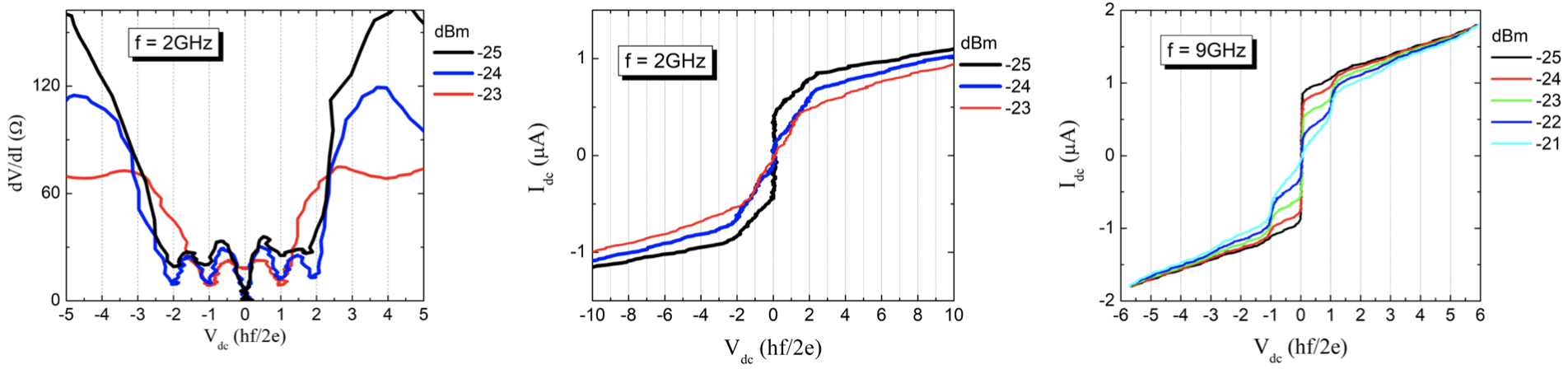}
\caption{\label{fig:device2} \textbf{a} The differential resistance at 2 GHz for a few microwave power levels. At this low frequency, both even and odd Shapiro steps are seen. \textbf{b} The corresponding I-V curves at 2 GHz.
\textbf{c} The I-V curves at 9GHz. The even Shapiro steps are suppressed, as shown in the differential resistance in Fig.~4 of the main text. }
\end{figure}

In Fig.~\ref{fig:device2} we present additional measurements of $dV/dI$ and VI curves at zero magnetic field. In Fig.~\ref{fig:device2}a, we show the differential resistance at 2 GHz for various microwave powers. We see that steps $0,~\pm1,~\pm2$ are clearly observed before dissipative effects wash out higher steps. The measured VI curves are shown in Fig.~\ref{fig:device2}b, where the steps are not easily resolved with the naked eye (hence, the need for $dV/dI$ measurements). Fig.~\ref{fig:device2}c shows VI curves at 9 GHz, showing a large first steps, the clear suppression of the second step, and a weak third step.

\end{document}